\documentclass[vecphys]{svmult}

\usepackage{makeidx}         % allows index generation
\usepackage[bottom]{footmisc}% places footnotes at page bottom

\usepackage{psfig}
\usepackage[dvips]{graphicx}
\usepackage{natbib}  %OK to leave this on all the time: bibstyle after 

\def\xmm{{\sl XMM-Newton}}
\def\asca{{\sl ASCA}}
\def\xte{{\sl RXTE}}

\def\exo{{\sl EXOSAT}}

\def\mcg6{{MCG-6-30-15}}

\def\mch{M$\rm^{c}$Hardy\,}

\def\ecs{ergs cm$^{-2}$ s$^{-1}$~}
\def\msun{$M_{\odot}$}
\def\mbh{{$M_{BH}$}}
\def\me{{$\dot{m}_{E}$}}

\def\lbol{{$L_{bol}$}}
\def\tb{{$T_{B}$}}
\def\td{{$T_{d}$}}
\def\rb{{$R_{BLR}$}}
\def\ltsim{\mathrel{\hbox{\rlap{\hbox{\lower4pt\hbox{$\sim$}}}\hbox{$<$}}}}
\def\gtsim{\mathrel{\hbox{\rlap{\hbox{\lower4pt\hbox{$\sim$}}}\hbox{$>$}}}}

\makeindex             % used for the subject index

\begin{document}
\title*{X-Ray Variability of AGN and Relationship to Galactic Black
Hole Binary Systems }
\titlerunning{AGN and Binary X-ray Variability}
\author{Ian M$\rm^{c}$Hardy\inst{1}}
\institute{School of Physics and Astronomy, University of
Southampton, Southampton S017 1BJ, UK
\texttt{imh@astro.soton.ac.uk}}
\maketitle \abstract{

Over the last 12 years, AGN monitoring by 
\xte\, has revolutionised our understanding of the X-ray variability
of AGN, of the relationship between AGN and Galactic black hole X-ray
binaries (BHBs) and hence of the accretion process itself,
which fuels the emission in AGN and BHBs and is the major source
of power in the universe. In this paper I review our current
understanding of these topics.

I begin by considering whether AGN and BHBs show the same X-ray
spectral-timing `states' (e.g. low-flux, hard-spectrum or `hard' and
high-flux, soft-spectrum or `soft'). Observational selection effects
mean that most of the AGN which we have monitored will probably be `soft state'
objects, but AGN are found in the other BHB states, 
although possibly with different critical transition accretion rates.

I examine timescale scaling relationships between AGN and BHBs.  I
show that characteristic power spectral `bend' timescales, \tb, scale
approximately with black hole mass, $M_{BH}$, but inversely with accretion
rate, \me\, (in units of the Eddington accretion rate) probably
signifying that \tb\, arises at the inner edge of the accretion disc.
The relationship $T_{B} \propto M_{BH}/\dot{m}_{E}$ is a good fit, implying
that no other potential variable, e.g. black hole spin, varies
significantly.  Lags between hard and soft X-ray bands as a
function of Fourier timescale follow similar patterns in AGN and
BHBs.

I show how our improved understanding of X-ray variability enables us
to understand larger scale properties of AGN. For example, the width
of the $H_{\beta}$ optical emission line, $V$, scales as
$T_{B}^{1/4}$, providing a natural explanation of the observed small
black hole masses in Narrow Line Seyfert Galaxies; if $M_{BH}$ were
large then, as $T_{B} \propto M_{BH}/\dot{m}_{E}$, we would require
\me$>1$ to obtain narrow lines.

I note that the rms X-ray variability scales
linearly with flux in both AGN and BHBs, indicating that the amplitude of the
shorter timescale variations is modulated by that of the longer
timescale variations, ruling out simple shot-noise variability models.
Blazars follow approximately
the same pattern. The variations
may therefore arise in the accretion disc and propagate inwards until
they hit, and modulate, the X-ray emission region which, in the case
of blazars, lies in a relativistic jet.

Short timescale (weeks) optical variability arises
from reprocessing of X-rays in the accretion disc, providing
a diagnostic of X-ray source geometry.
On longer timescales, variations in the disc accretion rate may
dominate optical variations.

AGN X-ray monitoring has greatly increased our
understanding of the accretion process and there is a strong case for
continued monitoring with future observatories.

}
\section{Introduction}
\label{sec:1}

Understanding the relationship between AGN, which are powered by
accretion onto supermassive black holes, and the much smaller
Galactic black hole binary systems (BHBs) is currently one of the
major research areas in high energy astrophysics. Possible
similarities between AGN and BHBs have been mooted ever since the
late 70's and early 80's when it was first realised that they were
both black hole systems \cite[e.g][]{white84}. However
comparison of their X-ray variability properties provided the
first quantitative method for this comparison \cite{mch88}. More
recently considerable attention has been devoted to the jet
properties of black hole systems and a strong scaling has been
shown by means of comparing radio and X-ray luminosities
\cite{merloni03,falcke04}. However here I concentrate on the
considerable insight which can be gained regarding the scaling
between AGN and BHBs by studying their X-ray timing similarities.

I begin this review with a discussion of AGN `states', as we must be
sure, when comparing AGN with BHBs, that we are comparing like with
like. I then discuss characteristic X-ray variability timescales and
how they scale with mass and accretion rate.  I then show how larger
scale AGN properties, e.g. AGN optical permitted line width, are
related to the small scale accretion properties. I discuss models for
the origin of the variability and show how the variability of blazars
fits the same pattern as for Seyfert galaxies, showing that the source
of the X-rays is separated from the source of the variations, which may
be the same in all accreting objects.  The long AGN X-ray monitoring
programmes which were set up to enable us to compare AGN and BHB X-ray
variability properties have also allowed us, through correlated
monitoring programmes in other wavebands, to understand a little about
what drives the optical variability of AGN. I therefore
conclude this review with an examination of that topic.

\section{AGN X-ray Variability and AGN `states'}
\label{sec:states}

\subsection{States of Galactic Black Hole X-ray Binary Systems}

BHBs are found in a number of
`states', originally defined in terms of their medium energy (2-10
keV) X-ray flux and spectral hardness. These states\index{states} are discussed
fully elsewhere in this volume \cite{belloni09,fender09} but,
for completeness here, and at the risk of some
over-simplification, I briefly summarise the properties of the
most commonly found states, i.e. the `hard', `soft' and `VHS'
states.

In the `hard' state the medium energy X-ray flux is low and the
spectrum is hard. In the `soft' state, the medium energy X-ray flux is
high and the spectrum soft. The main spectral characteristic of the
very high state (VHS), sometimes also known as the high intermediate
state, is that the X-ray flux is very high. The spectrum is
intermediate in hardness between the hard and soft states. The
accretion rate rises as we go from the hard to soft to VHS states. The
X-ray timing properties of the different spectral states are also very
different, and quite characteristic and may, in fact, provide a better
state discriminant than the spectral properties.

X-ray variability is usually quantified in terms of the power spectral
density (PSD) of the X-ray light curve.  In Cyg~X-1 there is a large
amplitude of variability in the soft-intermediate and soft states and
the PSD is well described by a simple bending powerlaw.  At high
frequencies the power spectral slope is -2 or steeper (i.e. $P(\nu)
\propto \nu^{-\alpha}$ with $\alpha \geq 2$) bending, below a
characteristic frequency $\nu_{B}$ (or timescale $T_{B}$) to a slope
$\alpha \simeq 1$ \citep{cui97a,mch04}.  In most other BHBs, eg
GX~339-4 \cite{belloni05}, which typically differ from Cyg~X-1 in
reaching the soft state only during very high accretion rate transient
outbursts but then reaching a more pronounced soft state, the
variable component is swamped by a quiescent thermal component and so
is very hard to quantify. However the PSD of the weak variable
component, when measurable, is consistent with the shape seen for Cyg
X-1.

In the hard state the PSD can also be
approximated by bending powerlaws, and the high frequency bend is also
seen. In addition, a second bend to a slope of $\alpha= 0$ is seen
approximately 1.5 to 2 decades below the high frequency bend.  However
in the hard state, BHB PSDs are better described by a combination of
Lorentzian shaped components \citep{nowak00}, with prominent
components being located at approximately the bends in the bending
powerlaw approximation to the PSD. In the very high state and, indeed,
in any state apart from the soft state, the PSD can be described by
the sum of Lorentzian components, as in the hard state. However in the
very high state, the corresponding Lorentzian frequencies are higher
than in the hard state.

\subsection{Quantifying AGN States by Power Spectral Analysis}

AGN typically have X-ray spectra similar to those of hard state BHBs
(i.e. photon indices $\sim2$), and much harder than soft state
BHBs. [Note that, unless specifically mentioned to the contrary, by
`AGN' I mean an active galaxy, such as a Seyfert galaxy, where the
emission does not come from a relativistic jet. Relativistically
beamed AGN, i.e. blazars, are discussed in Sect.~\ref{sec:blazars}.]
Simple spectra do not provide a particularly direct means of state
comparison as, in soft state BHBs, the very hot accretion disc leads
to a large, soft spectrum, thermal component which dominates the
medium energy emission. In AGN the accretion disc is much cooler and
so does not affect the medium energy flux, even at high accretion
rates. However if one can disentangle the relative contribution of the
accretion disc from the total luminosity then AGN and BHB do seem to
occupy broadly similar regions of  disc-fraction/luminosity diagrams
\cite[e.g.][]{kording06agn,fender09}).
Nonetheless timing properties may provide a cleaner and simpler state
discriminant and means of comparing AGN with BHBs. However there are
considerable difficulties in measuring AGN PSDs over a wide enough
frequency range that the equivalent of the bends seen in BHB PSDs may
be found. Assuming to first order that system sizes, and hence most
relevant timescales, will scale with black hole mass, \mbh, then as
$\nu_{B} \sim10$Hz for the BHB Cyg~X-1 in the high state \citep[\mbh
$\sim10-20$\msun;][]{herrero95,ziolkowski05}, we
might expect $\nu_{B} \sim10^{-7}$Hz for an AGN with \mbh\,=$10^{8}$
\msun. To measure such a bend requires a light curve stretching over a
number of years.

The first really detailed observations of AGN X-ray variability were
made by EXOSAT\index{EXOSAT}. In the case of NGC~4051\index{NGC~4051} \citep{lawrence87} and NGC~5506\index{NGC~5506}
\citep{mch87}, variability on timescales of less than $\sim$day was
shown to be scale-invariant, initially dashing hopes of finding some
characteristic timescale from which black hole masses could be deduced.  
However by combining the \exo\, observation of NGC~5506 with 
archival observations from a variety of satellites, it was possible to
make a PSD covering timescales from $\sim$years to$\sim$minutes.
Although poorly determined on long timescales, the PSD did roughly
resemble the PSD of a BHB and was sufficient to show that the scale
invariance which prevailed on timescales shorter than a day broke down
at timescales of around a few days to a week \citep{mch88,mch89}.
The bend timescale was
broadly consistent with linear scaling with mass.  But the long
timescale data were very poor. Subsequent archival searches produced a
measurement of a bend frequency in NGC~4151 \citep{papmch95} but again
the long timescale data were poorly sampled. However the long
timescale monitoring of AGN, on which the production of high quality
PSDs relies, was revolutionised in late 1995 with the launch of the
\xte\, which was specifically designed for rapid slewing, allowing for
monitoring on a variety of timescales from hours to years. As a result
a number of groups began monitoring AGN. Long timescale PSDs of good
quality were made and reliable bend frequencies were measured for a
number of AGN \citep[eg][]{mchxte98,ednan99}.

\subsubsection{Power Spectral analysis of irregularly sampled data}

Determining the shape of AGN PSDs from sampling that is usually
discrete, and irregular is not trivial. Fourier analysis of the raw
observational data will result in the true underlying PSD being
distorted by the window function of the sampling pattern.  The
non-continuous nature of most monitoring observations means that high
frequency power is not properly sampled, which results in that power
being `aliased', or reflected, back to lower frequencies. Also, as the
overall light curves have a finite length, low frequency power is not
properly measured and so additional spurious power leaks into the PSD
from low frequencies. This latter effect is known as red noise
leak. The shape of the true underlying PSD can, however, be estimated
by modelling. The procedure which we use is called {\sc PSRESP\index{PSRESP}}
\cite{uttley02}. This procedure builds on earlier work (the `response'
method, \cite{done92}) and other previous authors have also used a similar
method to analyse AGN X-ray variability \cite{green93}. In {\sc
PSRESP} we assume a model form for the underlying PSD and, using a now standard
prescription \cite{timmer95}, a light curve is then simulated. This
light curve is folded through the real observational sampling pattern
and a `dirty' simulated PSD is produced. This procedure is repeated
many times to establish an average model dirty PSD and to establish
its errors.  Observations from a number of different observatories (e.g.
\xte\, \xmm\, \asca) can be modeled simultaneously. The real,
observed, dirty PSD is then compared with the dirty model average and
the model parameters are adjusted until the best fit is obtained.

Using {\sc PSRESP} we have determined the overall PSD shapes of a
number of AGN, covering timescales from years to tens of seconds.
\citep[e.g.][]{uttley02,mch04,mch05,uttleymch05}. In only one case
(Akn~564, described below) do we find evidence for more than one
bend in the PSD. In all other cases we find only one bend, from
$\alpha=2$ at high frequencies to $\alpha=1$ at lower frequencies.
In NGC~4051\index{NGC~4051} \citep{mch04} and MCG-6-30-15\index{MCG-6-30-15} \citep{mch05} the slope
of $\alpha=1$ continues for at least 4 decades below the upper
bend and in a number of other AGN (e.g. NGC~3227\index{NGC~3227}, \cite{uttleymch05})
the slope can be traced for at least 3 decades. In such cases we
can be reasonably sure that a hard state PSD is ruled out and a
soft state PSD is required. In other cases there may not be
sufficient low frequency data to distinguish between hard and soft
states but in all cases a soft state PSD is consistent with the
data.

The lack of hard state AGN PSDs may, however, just be a selection
effect. Most of the AGN that have recently been monitored are X-ray
bright and have moderate accretion rates (typically above a few per
cent of the Eddington accretion rate, $\dot{m}_{E}$). In BHBs,
whenever the accretion rate is below $\sim0.02~\dot{m}_{E}$, the
sources are found in the hard state. Although hard states are also
found above $\sim0.02~\dot{m}_{E}$, no soft states are found below that
accretion rate. Thus the finding that almost all AGN PSDs may be soft
state PSDs is consistent with the expectation based on their accretion
rates. It is also possible that the transition accretion rate between
the hard and soft states is different in AGN. Soft states are
associated with an optically thick accretion disc reaching close in
to the black hole. In the lower temperature discs around more massive
black holes (i.e. AGN) it is possible that the optically thick part
of the accretion disc might survive without evaporation, at the same
accretion rate, at smaller radii than in the hotter discs around
BHBs. Thus the transition accretion rate might be lower around
AGN. However the data do not yet allow us to test this possibility.

There had been a tentative suggestion that there might be a second
bend, at low frequencies, in the PSD of NGC~3783\index{NGC~3783} and hence that it
might be a hard state system \citep{markowitz03a}. As NGC~3783 does
not have a particularly low accretion rate
($\sim0.07~\dot{m}_{E}$), this suggestion was a little puzzling.
However our further observations \citep{summons07} show that, in
fact, NGC~3783 is another soft state AGN.

\subsection{The Unusual case of Akn~564\index{Akn~564} - A very high state AGN}

\begin{figure}
\includegraphics[height=8.5cm,,width=8.5cm,angle=0]{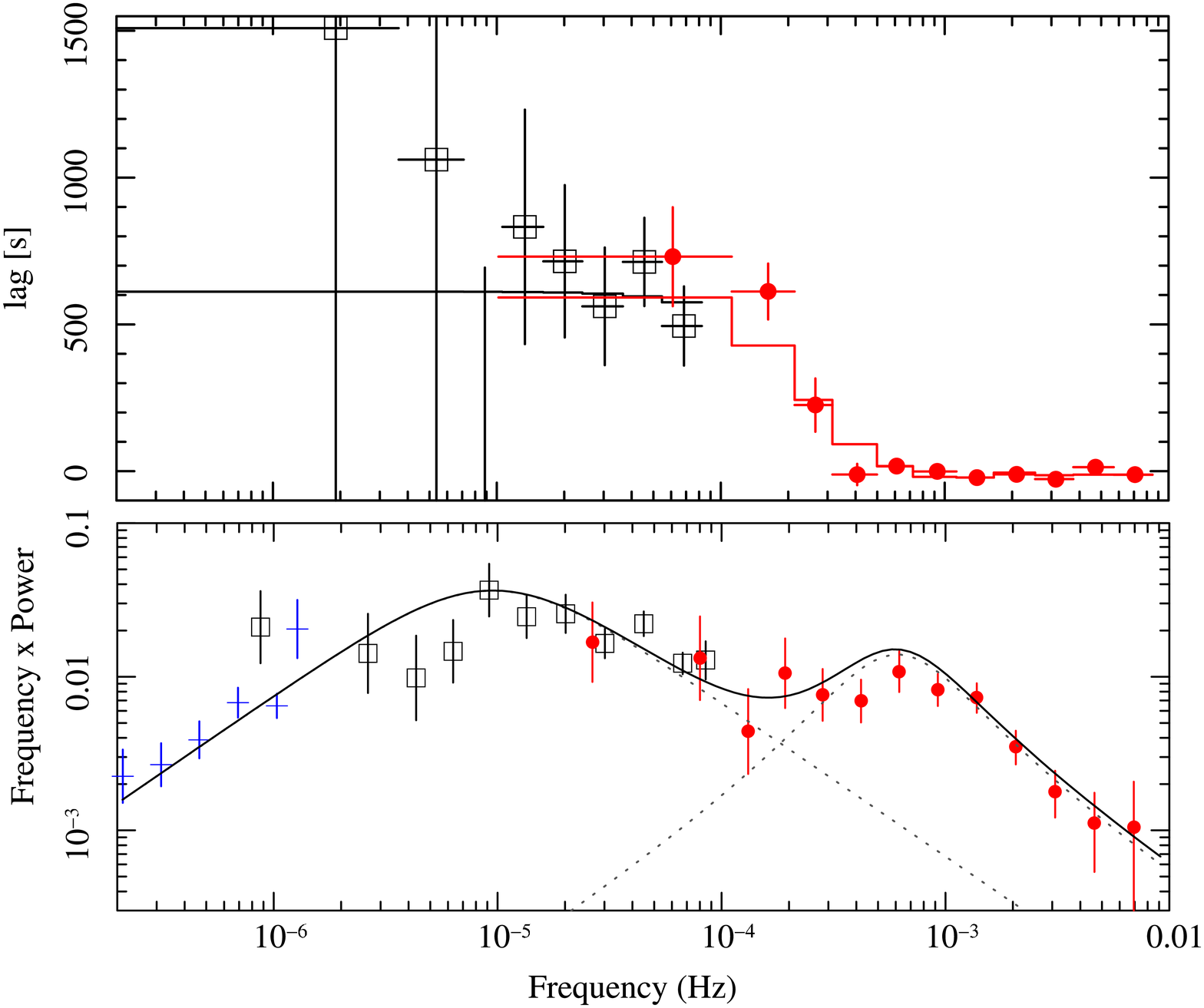}
\caption{Top panel: Lag between the hard (2-10 keV) and soft (0.5-2
  keV) X-ray bands as a function of Fourier frequency, for Akn~564\index{Akn~564}. A
  positive lag means that the hard band lags the soft. Above
  $10^{-3}$Hz the lags become slightly negative ($-11 \pm4.3$s), i.e. the
  soft band lags the hard band.
Bottom panel:
  PSD of Akn~564 showing a fit to the two-Lorentzian model
  \citep{mch07b}. Note how the cross-over frequency between the two
  Lorentzians corresponds to the frequency at which the lags rapidly
  change.
}
\label{fig:564}
\end{figure}

The finding of a VHS PSD in a high accretion rate AGN would
significantly strengthen the growing link between AGN and BHBs.
Arakelian~564 is one of the highest accretion rate AGN known
($\dot{m}_{E}\sim 1$) and is the only AGN for which there is presently
good evidence that its PSD contains more than one bend. From
observations with \xte, evidence for a low frequency bend was
presented \cite{pounds01} and evidence for a second bend, at high
frequencies, was also found from analysis of ASCA\index{ASCA} observations
\cite{papadakis02}. The combined \xte\ and \asca\ PSD was interpreted
as a hard state PSD \cite{papadakis02} although it was noted that the frequency
of the high frequency bend did not scale linearly with mass to that
of the hard state of the archetypal BHB Cyg~X-1; the frequency
difference is too small.  However proper modeling of the combined
long and short timescale PSD was hampered by the gaps which occur in
the \asca\ light curves, and hence in the PSD, at the orbital period
($\sim5460$s). We therefore made a 100ksec continuous observation
with \xmm\, to fill that gap in the PSD and the resulting overall
\xmm\, \asca\, and \xte\, PSD is shown in Fig.~\ref{fig:564}, bottom
panel.
Although the PSD of Akn~564 can be fitted with a doubly-bending
powerlaw as well as with two Lorentzian\index{Lorentzian} components, strong support for
the Lorentzian interpretation is given by the spectrum of the lags
between the hard and soft X-ray bands. It has long been noted in both
BHBs \cite[e.g.][]{nowak99} and AGN \cite[e.g.][]{papadakis01, mch04}
that the X-ray emission in the harder energy bands usually lags behind
that in the softer bands and that the lag increases with increasing
energy separation between the bands.  However if we split the
light curves into components of different Fourier frequency we can
measure the lag as a function of Fourier frequency, i.e. we can
determine the lag spectrum. Although I do not discuss it here, the
degree of correlation between light curves is known as the
coherence which, together with the lags, can provide a very useful
diagnostic of the source geometry and emission mechanism.
The calculation of lags and coherence is known as
cross-spectral analysis and a full description of the technique and it
applications is given elsewhere \cite{vaughan97}.

In BHBs where the PSD is well described by the sum of Lorentzian
components, i.e. in every state apart from the soft state, it is
found that the lags are reasonably constant over the frequency range
where the PSD is dominated by any one Lorentzian component, but the
lags change abruptly at the frequencies at which the PSD changes from
one Lorentzian component to the next.  In most AGN the PSD and lag
spectra are not defined well enough to distinguish different
Lorentzian components but for Akn~564 we can measure the lag spectrum
(Fig.~\ref{fig:564}, top panel). We see that the lag changes rapidly
from one fairly constant level to another at the frequency where the
PSD changes from one Lorentzian to another.
Given the extremely high accretion
rate of Akn~564, we interpret these observations as strongly
supporting the VHS, rather than hard state, interpretation.
We also note that, at the highest frequencies (above $10^{-3}$Hz), the
lags become slightly negative, i.e. the soft band lags
the hard band. One possible explanation is that, at these frequencies,
which may come from the very innermost part of the accretion disc,
the soft X-rays arise mainly from reprocessing of harder X-rays by the
accretion disc (see Sect.~\ref{sec:lags}).

\begin{figure}
\includegraphics[angle=270,width=8.5cm]{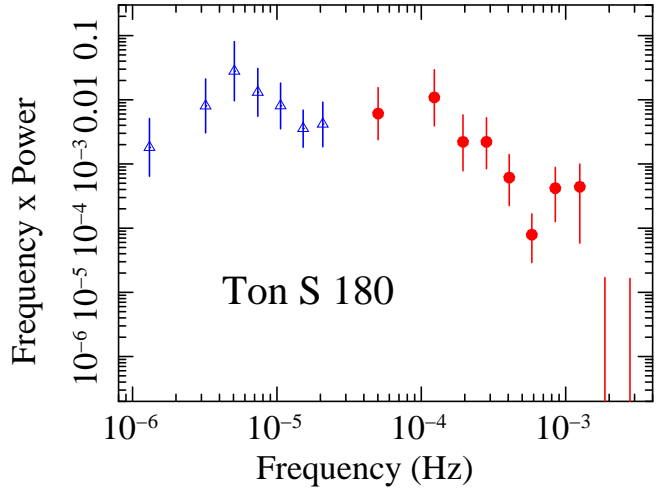}
\caption{Combined \xte\ and \asca\ PSD of Ton~S180, unfolded from the
  distortions introduced by the sampling pattern, assuming a simple
  model of a powerlaw with a bend at high frequencies. Two large bumps
are seen in the residuals which can be fitted equally well with two
Lorentzian-shaped components (Summons et al., in preparation)}
\label{fig:tons_psd}
\end{figure}

{\bf Ton~S180\index{Ton~S180}~} In our recent PSD survey of 32 well observed AGN
(Summons et al in preparation) the only AGN besides Akn~564 for
which there are reasonable hints that a double-Lorentzian, rather
than a single bending powerlaw model, is required is Ton~S180
(Fig~\ref{fig:tons_psd}). Interestingly Ton~S180 has, like Akn~564,
an accretion rate close to the Eddington limit and so may very
well be another VHS system.
As with Ark~564, we have calculated the lag spectrum but
the data are not as good as for Ark~564. The lags are, however,
consistent with a VHS interpretation, as in Ark~564.

\subsection{Periodicities in AGN}

\begin{figure}
\includegraphics[angle=0,width=8.5cm,height=7cm]{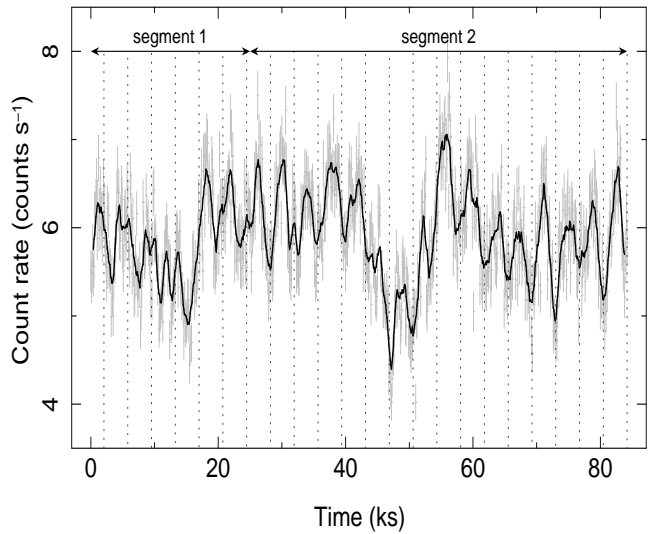}
\caption{\xmm\, 0.3-10 keV light curve of RE~J1034+396\index{RE~J1034+396}
  \citep{gierlinski08_qpo}. The thick black line represents the
  running average over 9 bins. The vertical dotted lines show the
  expected times of minima obtained from folding segment 2 with the
  period 3733s.
}
\label{fig:qpo}
\end{figure}

Although broad Lorentzian features can be fitted to the PSDs of Akn~564
and Ton~S180 there have been, until very recently, no believable
detections of highly coherent periodicities in AGN. However one
periodicity of high quality factor ($Q>16$) has now been found in the
NLS1 RE~J1034+396\index{RE~J1034+396} (Fig~\ref{fig:qpo}, from \cite{gierlinski08_qpo}). The
periodicity does change in character during the observation with
slight shifts in times of minimum and variations in amplitude so it is
very similar to the quasi-periodic oscillations (QPOs) seen in BHBs
\citep[e.g.][]{remillard06}.
 
The reason that a narrow QPO\index{QPO} is found in RE~J1034+396 is not known. It
has a high accretion rate, but so too does Akn~564 and Ton~S180. It has
been suggested that perhaps the phenomenon is transient as in BHBs
\cite{gierlinski08_qpo} and the observers were just very
lucky. Alternatively it may have something to do with the overall
spectral shape of RE~J1034+396 which is unique, even for an NLS1,
peaking in the far UV.  Whatever the reason, this
observation is very important as it shows that the X-ray variability
similarities between AGN and BHB extend to almost all known
phenomena. It also enables us to estimate the black hole mass.  In
BHBs there is a tentative relationship between the frequency of the
high frequency QPOs and the mass, $f_{0} =931 (M/M_{\odot})$Hz
\citep[e.g.][]{remillard06} where $f_{0}$ is the fundamental frequency
of the pair of high frequency QPOs whose frequency ratio is
$2f:3f$. In RE~J1034+396 then (and depending a little on which of the 2 high
frequency QPOs we think the observed periodicity is) we can estimate a
mass around $10^{7}~M_{\odot}$.

\section{Scaling characteristic timescales with mass and accretion
rate}

The early suggestions of a scaling\index{scaling} of \tb\, with \mbh\,
\citep{mch89,papmch95} were supported by the first results from the
\xte\, monitoring programmes
\cite[e.g.][]{mchxte98,ednan99,uttley02,markowitz03a}. However once
\tb\, had been measured in more than just a handful of AGN
it was clear that there was much more spread in
the \tb\, vs. \mbh\, relationship than could be accounted for just by
observational error. Interestingly it was noted that for a given \mbh, the
narrow line Seyfert 1 galaxies (NLS1s) had shorter values of
\tb\, than broad line AGN and so it was suggested that accretion rate,
as well as \mbh, might also affect the value of \tb\,
\citep{mch04,mch05,uttleymch05}. The reason for the dependence on
accretion rate was not clear but one possibility was that, if \tb\, is
somehow associated with the inner edge of the optically thick part of
the accretion disc, then if that edge moves closer to the black hole
when the accretion rate rises \citep{esin98}, then \tb\, would
decrease. Also, the increase in accretion rate would lead to an
increase in luminosity and hence in an increase in the inner radius of
the broad line region (\rb), and therefore in a decrease of the
typical velocities, or widths, associated with the emission lines
\cite{mch05}.

\subsection{Fitting the timing scaling relationship for soft-state objects}

In order to properly quantify the dependence of \tb\, on \mbh\, and \me,
and motivated by the approximate linear relationship between \tb\,
and \me\, and the approximate inverse relationship between \tb\, and
\me\, we hypothesised that

\[ log~T_{B}=A log~M_{BH} -B logL_{bol} +C \]

\noindent where $L_{bol}$ is the bolometric luminosity, and performed a
simple 3D parameter grid-search to determine the values of the
parameters $A, B$ and $C$. As the AGN under consideration were, in
all cases where the PSD was well defined, soft-state objects, then
\me $\sim L_{bol}/L_{Edd}$. However we preferred to fit to \lbol\,
rather than  \me\, as \lbol\,
is an observable, rather than a derived, quantity. The best
fit to a sample of 10 AGN was $A=2.17^{+0.32}_{-0.25}$,
$B=0.90^{+0.3}_{-0.2}$ and $C=-2.42^{+0.22}_{-0.25}$ \citep{mch06}.

\begin{figure}
\includegraphics[angle=270,width=8.5cm]{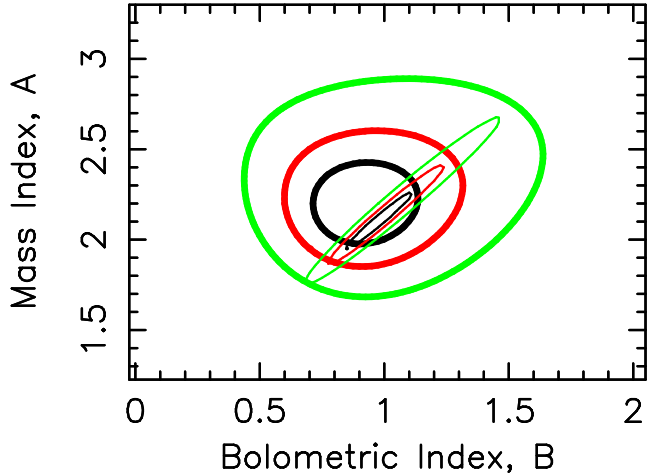}
\caption{Assuming a relationship between the PSD bend timescale, \tb\,
the black hole mass, \mbh\, and the bolometric luminosity, $L_{bol}$
of the form $Log\,T_{B}= A Log\,M_{BH} - B Log\,L_{bol} + C$, where $C$
is a constant, we present here the 68\% (black), 90\% (red) and 95\%
(green) confidence contours for $A$ and $B$. The thick contours are
derived from a fit to 10 AGN only. The thin contours also include the
BHBs Cyg~X-1 and GRS~1915+105 in their soft (radio
quiet) states. Note how the contours overlap completely, even at the
68\% confidence level.
}
\label{fig:AB}
\end{figure}

In order to determine whether the same scaling\index{scaling} extends down to
BHBs we included two bright BHBs Cyg~X-1(\index{Cyg~X-1} and GRS~1915+105\index{GRS~1915+105}) in
radio-quiet states where, for proper comparison with the AGN,
their high frequency PSDs are well described by the same cut-off,
or bending power-law model which best describes AGN and where
broad band X-ray flux provides a good measurement of bolometric
luminosity. For Cyg~X-1, we combined measurements of \tb\,
\citep{axelsson05} with simultaneous measurement of the bolometric
luminosity \citep{wilms06} over a range of luminosities. For GRS~1915+105, we measured an average \tb\, from the original X-ray data,
and determined \lbol\, from the published fluxes \citep{trudolyubov01}
and generally accepted distance (11 kpc). For the combined fit to
the 10 AGN and the two BHBs, we find complete consistency with the
fit to the AGN on their own, i.e. $A=2.1\pm{0.15}$,
$B=0.98\pm{0.15}$ and $C=-2.32\pm{0.2}$. The confidence contours
for the fit to the AGN on their own and to the combined AGN and
BHB sample are shown in Fig~\ref{fig:AB}. We can see that, even at
the 1$\sigma$ (i.e. 68\% confidence) level, the contours for the
AGN on their own, and for the combined AGN and BHB sample,
completely overlap (as do the offset constants, C). Thus we answer
a long-standing question and show, using a self-consistently
derived set of AGN and BHB timing data, that over a range of
$\sim10^{8}$ in mass and $\sim10^{3}$ in accretion rate, AGN
behave just like scaled-up BHBs. Assuming $\dot{m}_{E}=
L_{bol}/L_{Edd}$, then \tb\, $\propto M_{BH}^{1.12}/\dot{m}_{E}^{0.98}$.

\begin{figure}
\includegraphics[angle=270,width=8.5cm]{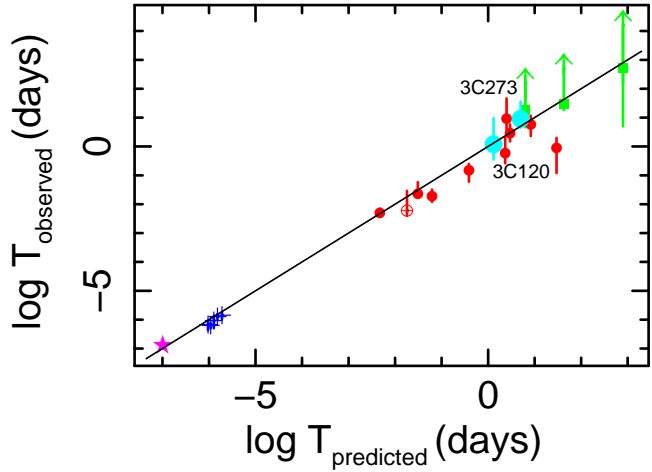}
\caption{Log of the observed vs. the predicted PSD bend timescales (in
units of days), using the best-fit relationship derived from
Fig.~\ref{fig:AB}. The star on the lower left is GRS~1915+105, the crosses
are Cyg~X-1 and the  circles are the AGN which were included in the
fit. The three squares, whose upper bend timescales are
unbounded, are AGN which were not included in the fit. However it can
be seen that the measured bend values are, nonetheless, consistent
with the fit from the AGN with well defined PSD bend
timescales. Similarly we
also plot the blazar 3C~273 (\mch~et al, in preparation, and
Sect.~\ref{sec:blazars} here) and the radio galaxy 3C~120 
(\cite{summonsphd} and Marshall et al, in preparation), which were also not included in the fit.
}
\label{fig:predict}
\end{figure}

We can see how well the scaling relationship fits the data by
comparing the observed bend timescales with the values predicted by
the best fit parameters (Fig~\ref{fig:predict}). The only noticeable
outlier is NGC~5506\index{NGC~5506} (at $log T_{predicted} = +1.3$). However we note
that the black hole mass estimate for NGC~5506  ($10^{8}$\msun) which we used
\cite{mch06} is based on the width of the [OIII]
lines, which is only a secondary mass estimator. The resultant implied
accretion rate (2.5\% \me) is surprisingly low given its recent
classification, from the width of the IR $P_{\beta}$ line, as an
obscured NLS1 \citep{nagar02_5506}. However using the recently
measured width of its stellar absorption lines \citep{gu06}, a smaller
mass (few $\times 10^{6}$\msun) is derived. With that lower mass,
NGC~5506 lies much closer to the best-fit line.

\subsection{Constraint on black hole spin\index{black hole spin}}
In the fits discussed above, the best-fit reduced $\chi^{2}$ was very
close to unity. These fits were performed without introducing any
additional systematic error into the fit, but purely by using the best
estimates of the observational errors. In many such fits an unknown
additional error has to be introduced to bring the reduced $\chi^{2}$
down to unity, but we do not require any additional source of
uncertainty. The implication is that no other parameter, which we do
not include in the fit, varies greatly from object to object. One
parameter which could conceivably affect \tb\, is black hole spin. In
faster spinning black holes, the innermost stable orbit is closer to
the black hole.  Thus if \tb\, is related to the radius of the inner
edge of the accretion disc, and if that edge reaches right up to the
last stable orbit, then \tb\, might be lower in faster spinning black
holes.  The implication of the good quality of the present fit is that
spin does not vary greatly from object to object. Thus if some X-ray
emitting black holes are considered, from the width of their Fe
$K_{\alpha}$ emission lines, to be maximally spinning
\citep{fabian05}, then probably they all are.

\subsection{Scaling relationship for hard-state objects}

An inverse dependence of a characteristic timescale on accretion rate
had been noted elsewhere \cite{migliari05}. There it was noted that the
frequency of the highest frequency lorentzian component, $\nu_{high}$,
in the hard-state PSD of the BHB GX~339-4, varied with radio
luminosity, $L_{R}$, approximately as $\nu_{high} \propto
L_{R}^{1.4}$. In standard jet models \cite[e.g.][]{blandford79} the
total power in the jet, $L_{J}, \propto L_{R}^{1.4}$. Assuming that
$L_{J} \propto \dot{m}_{E}$, then it is expected that
$\nu_{high} \propto \dot{m}_{E}$ \cite{migliari05}.

A similar scaling\index{scaling} of characteristic timescale with accretion rate is
seen in hard state observations of the BHB XTE~J1550-564\index{XTE~J1550-564}.  In their
Fig.3, Done et al \cite{done05} plot $\nu_{high}$ against a parameter
which they label `` \lbol''. However this parameter is derived by
simple linear transformation from X-ray flux, $S_{X}$.  From that same
Fig.3.  \cite{done05} it can be seen that, in 2002, when the source
was in a steady hard state,
$\nu_{high} \propto S_{X}^{0.45 \pm0.05}$.  In hard
state, jet dominated, systems, $S_{X} \propto \dot{m}_{E}^{2}$
\citep{markoff01}.  Thus for the 2002, hard state, $\nu_{high} \propto
\dot{m}_{E}^{0.9\pm0.1}$, i.e. the same dependence of timescale on
accretion rate in an individual object that was previously found for
the sample of AGN and BHBs.

In a comprehensive study it has been shown \cite{kording07a}, using radio
luminosity as a tracer of accretion rate in hard state objects
\citep{kording06}, that hard state black hole systems and neutron
stars also follow the same dependence of characteristic timescale
with \mbh\, and \me\, that was derived earlier \cite{mch06}, which thus
seems to be characteristic of the majority of accreting objects.

\subsection{RMS variability as a mass diagnostic}

From observations which are too sparse or too short to allow a well
defined PSD to be produced, it is possible to estimate the bend
timescale if we assume a particular, fixed, power spectral shape. The
variance in a light curve, $\sigma^{2}$, equals the integral of the
power spectrum over the frequency range sampled by the light curve. Thus
if we assume a fixed PSD shape, i.e.  typically a fixed normalisation,
low frequency slope and high frequency slope, then we can deduce the
bend frequency from the observed variance in the light curve
\cite[e.g.][]{papadakis04}.  This technique has been used to
demonstrate a strong relationship between \tb\, and black hole mass
but the dependence of \tb\, on \me\, has not been clear
\citep{papadakis04,nikolajuk04,nikolajuk06}. The reason that the
dependence on \me\, has not been clear in these analyses is probably
that the samples under consideration did not contain a great range of
accretion rates, with few NLS1s involved, and also that the assumption
of a universal, fixed, PSD shape is not absolutely correct.  Small
differences in normalisation (e.g. compare NGC~3227 and
NGC~5506 \citep{uttleymch05}) or in PSD slopes (the low frequency slope in
MCG-6-30-15 is -0.8 whereas in NGC~4051 it is -1.0 \citep{mch05,mch04})
can easily move the derived bend frequency by half a decade, or more,
which is sufficient to blur any dependence on accretion rate. High
frequency PSD slope is also a function of energy, being flatter at
higher energies \cite[e.g.][]{mch04}, and thus great care must be taken
to ensure uniform energy coverage. Also, in the translation from
$\sigma^{2}$ to \tb, a hard-state PSD shape has generally been assumed
whereas we now know that a soft state is generally more likely.

Nonetheless, $\sigma^{2}$ does still provide a useful diagnostic of
variability and has been used to good effect in the study of the
evolution of variability properties with cosmic epoch. In an analysis
of a deep ROSAT observation it has been shown \cite{almaini00} that,
for a particular luminosity, the variance of high redshift ($z>0.5$)
QSOs was greater than that of low redshift QSOs.  A similar, though
more detailed, analysis has been performed \cite{papadakis08} using
the extensive \xmm\, observations of the Lockman hole which show that,
for the same luminosity, black hole mass is lower, but accretion rate
is higher, for the higher redshift QSOs, indicative of the growth of
black holes with cosmic epoch.

\subsection{High frequency scaling relationships}

Measurements of rms variability are usually carried out on
individual observations lasting for maybe a day or so. These
observations thus mainly sample the high frequency part of the
PSD. There have been related methods of quantifying variability
which also use the high frequency part of the PSD. 

In the late 1980s we used the amplitude of the normalised PSD (after
removing the Poisson noise level) at a standard frequency of
$10^{-4}$Hz (the `normalised variability amplitude', $NVA$,
\cite{mch88} and \cite{green93}). This frequency was chosen to be as
high as possible whilst avoiding the region dominated by Poisson noise
in typical EXOSAT\index{EXOSAT} observations.  At that time very few AGN black hole
masses were available and so $NVA$ had to be plotted against
luminosity as a proxy, and $NVA$ was seen to decrease with
increasing luminosity. Subsequently, broadly similar techniques were
used \cite{hayashida98, done05} to show that high frequency
variability timescales scaled approximately with mass.

Recently the PSDs of a number of hard-state BHBs have been studied and
it is found that the high frequency part of the PSD does not change
greatly with source flux \cite{gierlinski08_mass} although the lower
frequency part does change. It is  surmised that the
relatively unchanged high frequency tail part of the PSD may represent
some limit, such as the last stable orbit, for the black hole in
question. Nonetheless a plot of the amplitude of the PSD tail extrapolated
to 1Hz ($C_{M}$ - very similar to the $NVA$ discussed above) against
mass reveals no correlation with mass within the BHB sample
alone. However if a sample of AGN \cite{nikolajuk06} are
included, estimating $C_{M}$ from the variance by model fitting
assuming a hard state PSD, there is a correlation with mass.  As many
AGN are probably soft state systems, and as the range of accretion
rates in that AGN sample is small, further work is
required to determine whether the high frequency part of the PSD gives
a mass measurement, independent of any accretion rate variations.

Overall, measures of variability timescale based solely on
the high frequency part of the PSD can provide a reasonable mass
estimate for many AGN samples where there is not a great range in
accretion rate. Future work may show that they give an accretion-rate
independent measure of mass, but that is not yet confirmed.
These methods have the advantage of being
applicable to short observations. However measurement of the PSD
bend timescale is more robust to changes of PSD slope and
normalisation and provides a more sensitive diagnostic of mass and
accretion rate.

\section{Relationship between nuclear variability properties and
larger scale AGN properties}

One of the more important observable AGN parameters is the width of
their permitted optical emission lines, $V$. This linewidth is often
used as a means of classifying different types of AGN. It has been
shown that the most variable X-ray emitting AGN (where variability has
typically been quantified in terms of $\sigma_{rms}$), tend to have
the narrowest optical emission lines \cite[e.g.][]{turner99}. However
there is a good deal of spread in the relationship between
$\sigma_{rms}$ and $V$, and no quantitative physical explanation of
the relationship has yet been forthcoming.

\begin{figure}
\includegraphics[angle=270,width=8.5cm]{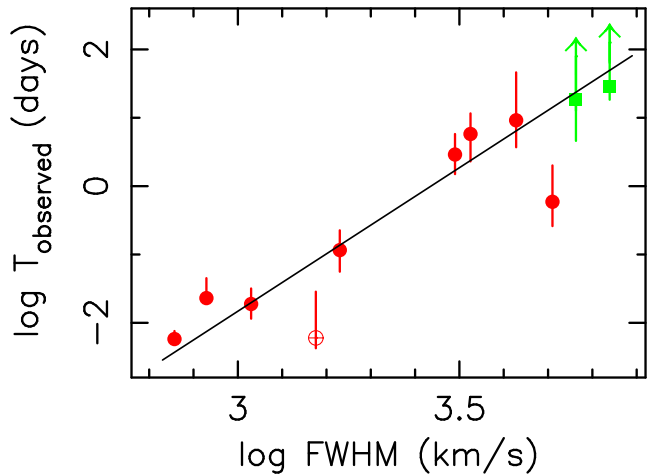}
\caption{Correlation of the observed PSD bend timescale vs. the FWHM
  of the $H_{\beta}$ optical emission line.
}
\label{fig:linewidth}
\end{figure}

In Fig~\ref{fig:linewidth} we show a plot of linewidth, $V$,
against \tb\ \cite{mch06}. Here we note a remarkably tight
relationship. Parameterizing $log T_{B} = D\,log V + E$,
we found that $D=4.2^{+0.71}_{-0.56}$  \cite{mch06}. Such a tight
relationship should have a simple explanation and, indeed, such an
explanation can be derived from simple scaling relationships.

We assume that the linewidth arises from the Doppler motions of
neutral gas at the inner edge of the broad line region (BLR), and that
the BLR is excited by radiation from the central black hole.  If the
gas is in virial motion then $V^{2} \sim GM_{BH}/R_{BLR}$, where \rb\,
is the radius of that inner edge. We assume that the ionising
luminosity, $L \sim M_{BH}\,\dot{m}_{E}$.  For the locally optimised
condition for the production of emission lines by gas at the same
optimum density and ionisation state we expect $R_{BLR} \propto
L^{0.5}$ and measurements of the radius of the BLR confirm that
expectation, finding that $R_{BLR} \propto L^{0.518\pm0.039}$
\cite{bentz06}. Putting these simple scaling relationships together we
then expect that $V^{4} \sim M_{BH}/\dot{m}_{E}$. As it has already
been shown that \tb\, $\sim M_{BH}/\dot{m}_{E}$, then we naturally
explain the observation that \tb\, $\sim V^{4}$ and, by
self-consistency, we show that the basic assumptions that we have made
above about the BLR are probably correct.

We therefore have a strong link between small scale nuclear
accretion properties (X-ray variability) and larger scale
properties of the AGN (linewidth). The relationship shown in
Fig~\ref{fig:linewidth} is particularly useful as it is purely a
plot of two observational quantities, and does not rely on any
assumptions or derivations. Thus from a simple measurement of an
optical linewidth, we can make at least a first order prediction
of how the X-ray emission from a Seyfert galaxy should vary.

\section{Origin of \tb\,}

We have now measured, reasonably well, the way in which \tb\, varies
with \mbh\, and \me\, in all types of accreting systems. But what is the
physical origin of \tb\,?  On timescales shorter than \tb\,, the
variability power drops rapidly and so it is natural to associate \tb\,
with some cut-off, or edge, in the accreting system. One obvious
possibility, therefore, is the inner edge of the optically thick
accretion disc. However, unless we push the inner edge of the disc out
to much larger radii ($\gtsim 20 R_{G}$) than are considered
reasonable for soft state systems \citep{cabanac09}, the
simple dynamical timescale, \td, at the edge of the disc is much
shorter than the observed PSD bend timescale. Other possible
characteristic timescales are the thermal timescale, $T_{therm} =
T_{d}/\alpha$, where $\alpha$ is the viscosity parameter, typically
taken to be $\sim0.1$, and the viscous timescale, $T_{visc}=
T_{therm}/(H/R)^{2}$, where $H/R$ is the ratio of the scale height to
radius of the disc \citep{treves88}.

For any fixed black hole mass, the viscous timescale, as well as the
dynamical timescale, can be varied only by varying the inner disc
radius (assuming that the viscosity parameter does not
change). Measurements of the disc radius, usually by means of spectral
fitting to the disc spectrum (e.g. \cite{cabanac09}), show
that the inner disc radius is large at low accretion rates, and
decreases as the accretion rate rises. However once the accretion rate
exceeds a few percent of the Eddington accretion rate (i.e. typically in
soft states), the radius has typically decreased to a few
gravitational radii and it is hard to measure further changes with
increasing accretion rate.  Although such
measurements are prone to large uncertainties, nonetheless there
does not yet seem to be evidence for the large changes of
radius with accretion rate that would be required to explain the
spread of timescales, for a given black hole mass, that are
observed \cite[e.g.][]{mch05}.

The viscous timescale can, however, be changed, for the same radius,
by altering the scale height of the disc. If we assume that, when in
the soft state, the inner edge of the accretion disc reaches to the
innermost stable orbit at, say $1.23 R_{G}$ (the limit for a maximally
spinning Kerr black hole), and that the observed bend timescale is
actually a viscous timescale at that radius, then we can derive the
scale height of the disc ($H/R$). We then find that $H/R$ varies
between about 0.1 for \me\ $\sim 0.01$ to about unity at  \me\ $\sim
1$, which is a sensible range, broadly in line with theoretical
expectations. Thus \tb\ might well be associated with the viscous
timescale at the inner edge of the accretion disc: that edge
firstly moves in as the accretion rate rises in the hard state, and
then, once the edge has reached the innermost stable orbit, further
rises in accretion rate increase $H/R$.

\section{Origin of the Variations}

We have seen that AGN and BHBs have similar `states\index{states}' and display
similar patterns of variability. We have found scaling relationships
which enable us to link the timing properties of accreting objects of
all masses. We are beginning to have some idea of the origin of the
main characteristic timescale in AGN. But what is the
underlying origin of the variability?

\begin{figure}
\begin{minipage}{55mm}
\includegraphics[angle=0,width=5.6cm]{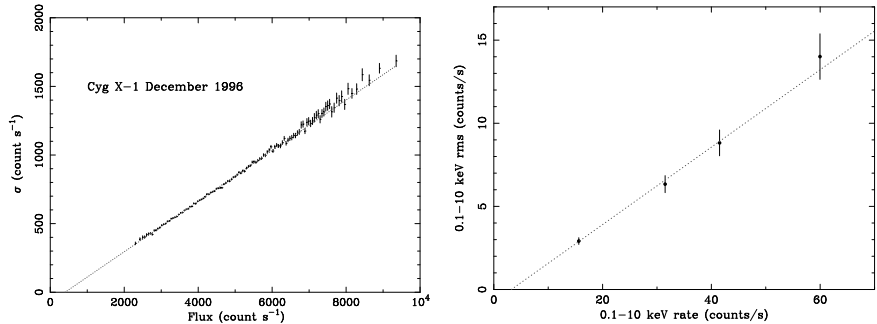}
\end{minipage}
\hspace*{2mm}
\begin{minipage}{55mm}
\includegraphics[angle=270,width=5.6cm]{chap08_4051_rmsflux.ps}
\end{minipage}
\caption{Left panel: rms-flux relationship for Cyg~X-1
  \citep{uttleymch01}. Right panel: rms-flux relationship for the
  Seyfert galaxy NGC~4051 \citep{mch04}
}
\label{fig:rmsflux}
\end{figure}

\subsection{Shot-noise models}

A model commonly-used to describe variability is the random 
shot noise\index{shot noise} model. In this model, light curves are made up of the
superposition of pulses, or shots, randomly distributed in time.
If all of the shots have the same decay timescale then we have
white noise. However if we allow for a range of decay timescales
then we can reproduce the bending powerlaw PSDs seen in BHBs and
AGN \citep{belloni90,lehto89,mch89}. Also, if more luminous
sources simply contain more shots, then the fractional variability
of more luminous sources will be less than that of less luminous
sources. This behaviour is generally seen but from the fact that
the normalised variability amplitude, $NVA$, decreases with
increasing luminosity less slowly than would be expected for
random, similarly sized, shots (i.e. we see $NVA \propto L^{-0.34}$
rather than the expected $NVA \propto L^{-0.5}$), it was concluded
that if the variability does arise from the summed emission from a
number of separate regions, then
there must be some correlation between those regions
\citep{green93}.

\subsection{The rms-flux relationship\index{rms-flux relationship}}

Although shot noise models can explain the shape of the
power spectra, they cannot explain the way in which the rms
variability of the light curves varies with flux. Following some
discussions regarding the benefits of calculating PSDs in absolute
or rms units, it was noted that the rms variability of the
light curves is directly proportional to the flux
\citep{uttleymch01}. This proportionality is now known as the
rms-flux relationship.  In Fig.~\ref{fig:rmsflux} we show the
rms-flux relationship for the BHB Cyg~X-1\index{Cyg~X-1} \citep{uttleymch01} and
for the Seyfert galaxy NGC~4051\index{NGC~4051} \citep{mch04}. This linear
relationship has now been found in all BHBs and AGN
for which the observations were sufficient to define it.
We incidentally note that the rms-flux relationship usually does not
go through the origin but has a positive intercept on the flux axis at
zero rms, indicative of a constant (e.g. thermal) component to the X-ray
flux. 

The important implication of the rms-flux relationship is that
when the long timescale Fourier components of the light curve,
which determine the overall average flux level, are large, the
short timescales components, which determine the rms variability,
are also large, and vice versa.  Thus different timescales know
about each other.

In simple additive shot noise models the shots are independent and
so the rms will be constant and independent of flux.  Simulations
show that a random distribution of similar shots will not
reproduce the extended low-flux periods seen in NGC~4051
\citep{guainazzi98, uttley99}, i.e. the short timescale variability
knowing that the average long timescale flux is very low. In
simple shot noise models at least one large shot will always
appear in the low flux periods. As the rms/flux relationship
applies, at least to BHBs, on whichever timescale one measures the
rms (e.g. 1s, 10s bins etc), then to make shot noise models produce
the rms-flux relationship, we would have to introduce some
mechanism which alters the shot amplitude on a large variety of
timescales. What that mechanism would be is unknown and hence we
are really no further forward in our understanding.

\subsection{Propogating fluctuation models}

An alternative, and simpler, explanation of the rms-flux relationship
is provided by the propagating fluctuation model, initially proposed
by Lyubarskii \cite{lyubarskii97}. In this model fractional variations
in mass accretion rate are produced on timescales longer than the
local viscous timescale so that longer timescale variations are
produced further out in the accretion disc. These variations then
propagate inwards.  If the low frequency variations then modulate the
amplitude of higher frequency variations produced closer in (as in
amplitude modulation in radio communications) then we produce a linear
rms-flux relationship\index{rms-flux relationship}.  Note that this relationship between
frequencies is multiplicative, not additive. If each inner frequency
is modulated then by the product of all lower frequency modulations
produced further out, then the rms-flux relationship will apply no
matter what frequency range one chooses to measure the rms over, or at
which lower frequency one chooses to measure the flux.

There are a number of implications when the time series is
produced by multiplying together a set of variations of different
frequencies. First let us consider the distribution of fluxes. If
a light curve is produced by the sum of many random sub-processes
then, by the central limit theorem, the fluxes will be normally
distributed, i.e. they will follow a Gaussian distribution.
Summative processes are linear \cite[e.g.][]{priestley82}, i.e.
multiplying the input (e.g. the accretion rate) by some constant
will multiply the output (e.g. the luminosity) by the same factor.
However if a light curve is produced by the product of many random
sub-processes, then the flux distribution will be lognormal (ie
the distribution of the log of the fluxes will be Gaussian). The
flux distributions of the well measured BHBs such as Cyg~X-1 do
indeed follow such a lognormal distribution. (A full discussion of
these topics is presented elsewhere \cite{uttley05}.)

\begin{figure}
\includegraphics[angle=0,width=11.5cm,height=8.5cm]{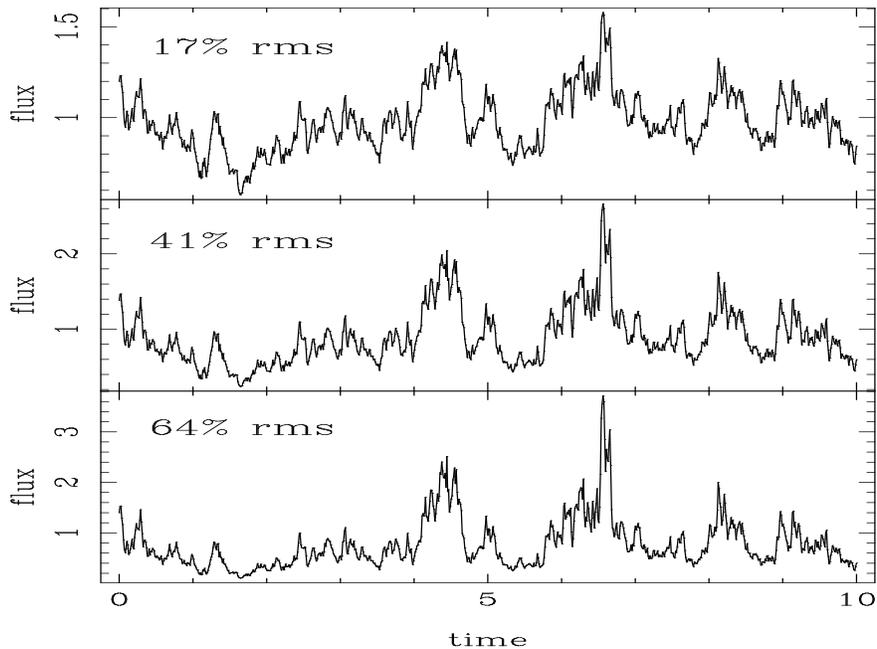}
\caption{Simulated exponential light curves \cite{uttley05}
showing increasing non-linearity as the rms variability increases.
}
\label{fig:explcs}
\end{figure}

If the observed light curve is actually the product of variations
on a variety of timescales, then it is straightforward to show
\citep{uttley05} that the observed light curve is the exponential
of an underlying linear light curve. The observed light curve is
therefore non-linear. As the non-linearity arises from the
coupling between variations on different timescales, being
ultimately driven by the lowest frequency variations (from the
outer part of the disc, in the Lyubarskii model), then the larger
the driving variations are, the greater will be the non-linearity.
In Fig.~\ref{fig:explcs} I show examples of light curves simulated
using the exponential formulation, with increasing rms variability
(from top to bottom) but with similar random number sequence used
in their generation. It can clearly be seen how the light curves
become more non-linear as the rms variability increases. The lower
rms variability light curves are typical of many ordinary Seyfert
galaxies whilst the higher rms, more non-linear, light curves are
typical of the NLS1s.

A qualitative model which can explain many of the spectral timing
observations of AGN and BHBs has been proposed 
\cite{churazov01, kotov01}. In this model
the 1/f power spectrum of variations which is produced
in the Lyubarskii model propagates inwards, perhaps in an
optically thin corona over the surface of the disc, until it hits
the X-ray emitting region, whose emission it modulates. A
numerical version of this model has been developed
\cite{arevalouttley06} which can reproduce many observed aspects
of X-ray spectral variability. The critical aspects of these
models is that the source of the variations is not the same as the
source of the X-rays.

\subsubsection{Lags and Coherence}
\label{sec:lags}

As well as providing an indication of X-ray state
(Sect.~\ref{sec:states}), lags and coherence are potentially very
powerful diagnostics of emission models and geometry. A standard
interpretation of the lag of the soft band by the hard band is given
by Comptonisation models. In these models the low energy seed photons
from the accretion disc are Compton-scattered up to higher energies by
a surrounding corona of very hot electrons. As more scatterings are
required to reach the higher energies, the higher energy X-rays will
be delayed relative to the lower energy photons.

\begin{figure}
\includegraphics[angle=270,width=10cm]{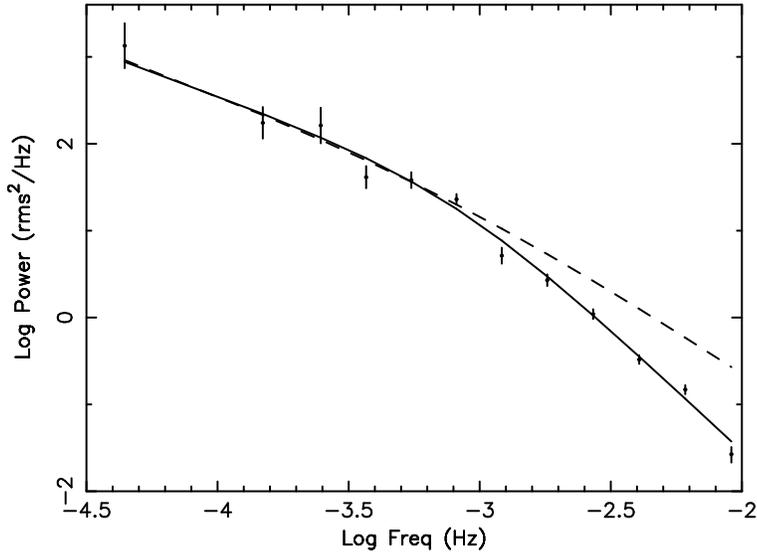}
\caption{PSDs of NGC~4051\index{NGC~4051} from \xmm\ observations. The data are from
  the 0.2-2 keV band and the solid line is the best fit to these
  data. The dashed line is the fit to the 2-10 keV PSD. Note that this
  PSD is flatter than that of the 0.2-2 keV PSD at higher
  frequencies. (The normalisation of the 2-10 keV PSD has been
  slightly increased to make it the same as that of the 0.2-2 keV PSD
  at low frequencies.)
}
\label{fig:4051psds}
\end{figure}

However there are other observed aspects of X-ray variability
which are not so easily explained by simple Comptonisation
scenarios, e.g. the observation that the slope of the PSD above
the bend is flatter for most AGN at higher energies (eg
Fig.~\ref{fig:4051psds}). If the variability is driven by
variations in the seed photons, one would expect high frequency
variability to be washed out by many scatterings and so would
expect the PSD to be steeper above the bend at higher frequencies.
We might retain the Comptonisation model if the high frequency
variability is driven by variations in the corona rather than in
variations in the seed photon flux. However the temperature and
optical depth of the corona would have to be carefully tuned so
that seed photons are raised to high energies by very few
scatterings and are then immediately able to leave the corona.

A perhaps more natural explanation of the lags, the coherence and the
energy-dependent PSD shapes can be provided by the propagating
fluctuations scenario if we introduce an X-ray emitting region with a
hardness gradient, such that higher energy X-rays are emitted,
preferentially, closer in towards the black hole \cite{kotov01}. The
geometry of the emission region is not critically defined but might
extend over the inner region of the accretion disc. The overall
emissivity at all energies would still rise with decreasing radius,
perhaps $\propto r^{-3}$, in accordance with the expectation for the
release of gravitational potential energy, but the proportion of
higher energy photons released would increase with decreasing
radius. If viewed from far out in the accretion disc, there would thus
be an emission-weighted centroid for each energy, and this centroid
would move to larger radius as the energy decreases. The observed lag
between the hard and soft bands would therefore simply represent the
time taken for the incoming perturbations to travel between the
centroids of the particular soft and hard energy bands under
consideration. 

This scenario can also explain why the lags vary with the Fourier
frequency of the perturbations.  The lowest frequency perturbations
will originate far beyond any X-ray emission region and so will
produce the largest possible lag. The lag will remain the same as we
move to higher frequencies until we reach frequencies which are
produced within the X-ray emission region.  Once we move inside the
X-ray emission region then the centroid of the remaining emission from
smaller radii (i.e. the emission that can still be modulated by the
perturbations from further out) will move to smaller radii. As the
radial profile of the X-ray emission will be more extended at lower
energies, the centroid of the lower energy emission will move inwards
more than the centroid of the higher energy emission. Therefore the lag will
decrease as we move to higher frequencies (e.g. NGC~4051, \cite{mch04}).

In Akn~564 we note (Fig~\ref{fig:564}) that, at the very highest
frequencies, the lags are slightly negative ($-11 \pm
4.3$s {\cite{mch07b}), i.e. the hard band leads. In the model discussed here
the very high frequencies are produced in the region where the X-ray
source is very hard. The main source of soft photons in this region
may therefore not be the intrinsic source spectrum but may come from
reprocessing of the intrinsic very hard spectrum by the accretion disc.
The lag may then represent approximately twice the light travel
time between the hard X-ray corona and reprocessing accretion
disc. 

By altering the emissivity profile of the disc it is possible to
produce lags which vary as a function of frequency either smoothly (as
in NGC~4051 or Mkn~335\index{Mkn~335}, e.g. Fig.~\ref{fig:335} \cite{arevalo08_335}) or
in a stepwise manner (e.g. Akn~564, \citep{mch07b}). The propagating
fluctuation model can simultaneously explain the energy-dependent
shape of the PSDs, thereby providing a simple and self-consistent
explanation of many spectral-timing properties of accreting systems.

\begin{figure}
\includegraphics[angle=270,width=11.5cm]{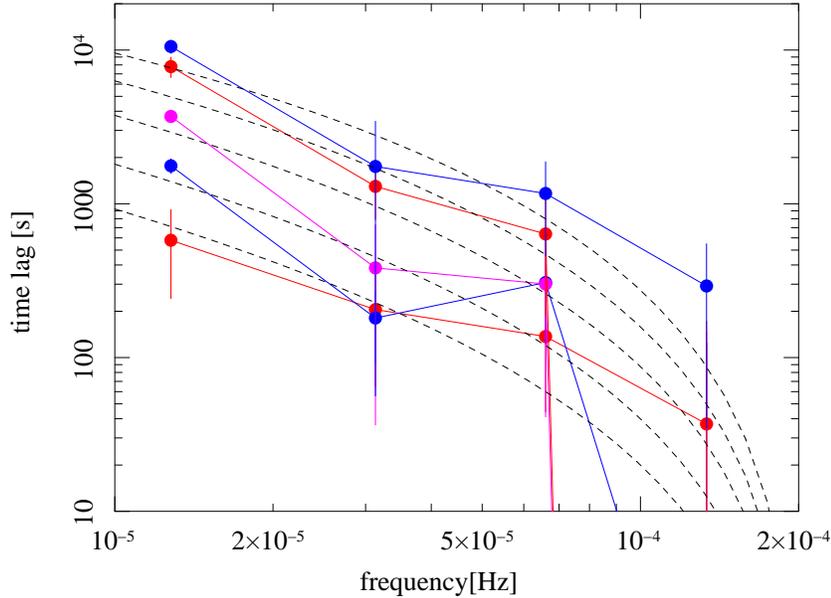}
\caption{The data points show the
time lag, as a function of Fourier frequency, between the
  0.2-0.4 keV band and the 0.4-0.6, 0.6-1, 1-2, 2-5 and 5-10 keV
  bands for Mkn~335. The lags increase as we move to larger
  energy separations and in all cases the hard band lags behind the soft band.
The faint dashed lines represent model fits to these data based upon
the propagating fluctuation scenario with a disc emissivity profile
(at least in the 0.2-0.4 keV band) $\propto r^{-3}$ and the inner edge
of the optically thick disc being truncated at $6R_{g}$ \citep{arevalo08_335}.
}
\label{fig:335}
\end{figure}

We note that the truncation of the optically thick disc leads to a
rapid drop in the lags at a frequency close to the bend in the PSD.
Although we do not yet have accurate lag measurements for many AGN,
the indications are that the lag drops follow the same mass (and
possibly accretion rate) scaling that is found for the PSD bend
timescales.

\section{Variability of Blazars}
\label{sec:blazars}

The radio through optical emission in blazars\index{blazars} like 3C~273 and 3C~279, ie
AGN with apparent superluminal radio components, is synchrotron
emission from a relativistic jet oriented at a small angle to the line
of sight. In 3C~273 the tight correlation between the synchrotron
component (e.g. IR) and X-rays, with the X-rays lagging the synchrotron
by $\sim$day \citep[e.g.][]{mch07a}, indicates that the X-ray emission
mechanism is almost certainly synchrotron self-Compton (SSC) emission
from the jet \citep[e.g.][]{sokolov04, mch07a}. A similar mechanism
probably applies in all other blazars.  Thus the X-ray emission region
(in a relativistic jet) and emission process are different to those of
Seyfert galaxies where the X-ray emission is not relativistically
beamed and the emission process is probably a thermal Compton emission
process. It is therefore not at all clear that we should expect
similar X-ray variability characteristics.

3C~273\index{3C~273} is one of the brightest AGN in the sky and, as such, has been
observed by all major observatories since the 1970s. We have collected
these data, including extensive observations with \xte, and the long
term light curve is shown in Fig~\ref{fig:273}a.  Using {\sc PSRESP},
in our normal manner, I have made the PSD which is shown, unfolded
from the distortions introduced by the sampling pattern, in
Fig~\ref{fig:273}b. It can be seen that this PSD is exactly like that
of a soft state Seyfert galaxy PSD.  The bend timescale is 10 days. I
have placed this observed timescale together with the predicted bend
timescale onto Fig.~\ref{fig:predict}. For comparison with most of the
other non-beamed AGN on that figure we use the black hole mass derived
from reverberation measurements by the same observers
\cite{peterson04}. It can be seen that 3C~273 fits the scaling
relationship between \mbh, \me\, and \tb\, as well as all the Seyfert
galaxies.

\begin{figure}
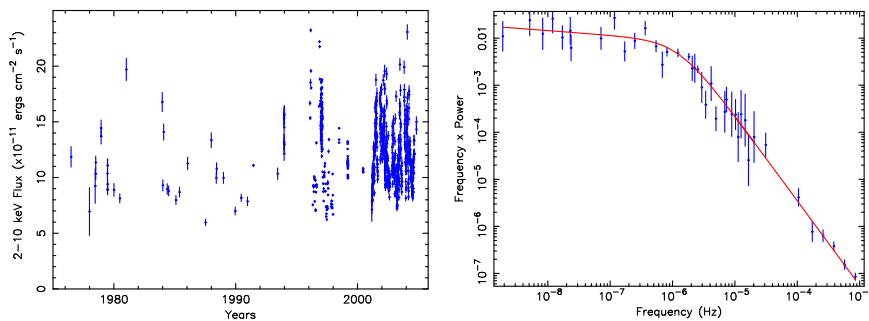

\begin{minipage}{55mm}
\includegraphics[angle=270,width=5.6cm]{chap08_all_lc_year_col.eps}
\end{minipage}
\hspace*{2mm}
\begin{minipage}{55mm}
\includegraphics[angle=270,width=5.6cm]{chap08_unfold_feb05_col.eps}
\end{minipage}
\caption{{\bf (a)} Left panel: Long term 2-10 keV X-ray lightcurve of
  3C~273. The intensive monitoring after 1996 is from \xte\,;   the
  earlier data are from the Ariel V SSI, Exosat and Ginga.
{\bf (b)} Right panel: Unfolded PSD of the blazar 3C~273, showing an excellent fit to
  the same bending powerlaw model that fits the Seyfert galaxies (\mch
  et al, in preparation)}
\label{fig:273}
\end{figure}

We note that there is no necessity to alter the observed bend
timescale by any relativistic time dilation factor to make the
observed bend timescale fit with the relationship derived for
Seyfert galaxies and BHBs. For a clock moving with the jet of
3C~273, that factor would be about 10. The implication, therefore,
is that the source of the variations - note, not the source of the
X-rays - lies outside the jet and is simply modulating the
emission from the jet. Coupled with the fact that the PSD of 3C~273
looks exactly like that of soft state Seyfert galaxies and BHBs,
this implication is then entirely consistent with Seyfert
galaxies, BHBs and blazars all suffering the same sort of
variations, e.g. variations propogating inwards through the disc.
The jet then would be seen as an extension of the corona which
dominates the emission in Seyfert galaxies.

\subsubsection{rms-flux relationship for 3C~279}
3C~279\index{3C~279} is an even more relativistically beamed blazar than 3C~273
\cite[e.g.][]{chatterjee08}. For both of these objects we have
calculated the rms-flux relationship\index{rms-flux relationship} from the \xte\, monitoring
observations. Both objects show a strong linear relationship
(e.g. Fig~\ref{fig:279}). This relationship provides strong
confirmation that the underlying process driving the variability in
blazars has the same multiplicative relationship between different
frequencies that is seen in BHBs and Seyfert galaxies. Coupled with
our observation that the PSD shape of 3C~273 is identical to that of
Seyfert galaxies, the simplest conclusion is that the process driving
the variability in all systems is the same, i.e. fluctuations
propagating inwards through the disc.

\begin{figure}
\includegraphics[angle=270,width=11.5cm]{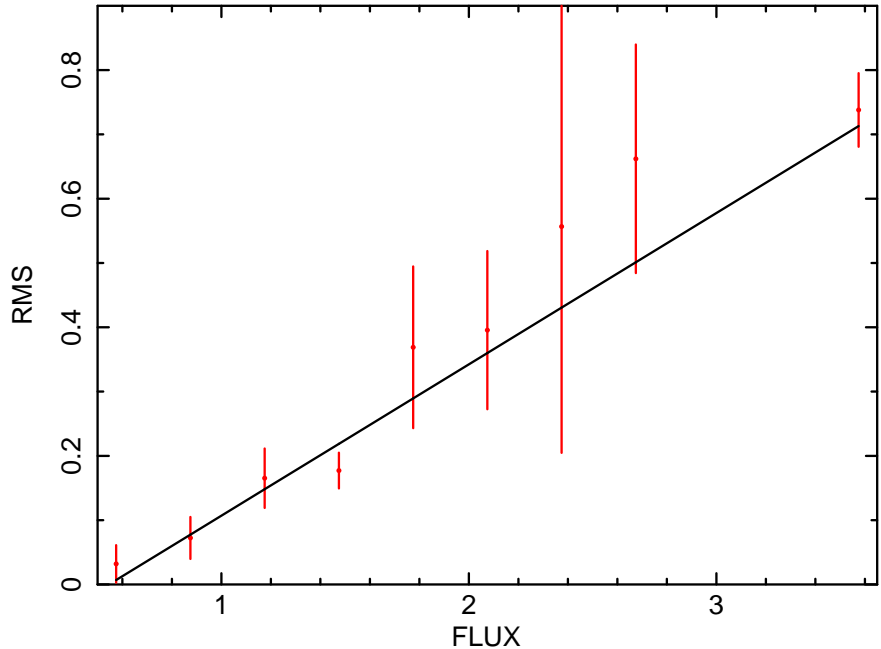}
\caption{RMS-Flux relationship for 3C~279 with flux and rms both being
  measured in the 2-10 keV band in units of $10^{-11}$ \ecs.
}
\label{fig:279}
\end{figure}

\section{X-ray/Optical Variability}

For many years the origin of the optical variability of AGN has been,
at best, a mystery and, at worst, a complete muddle. It has often been
suggested that optical variability arises from reprocessing of X-ray
variations but most early X-ray/optical studies found either weak
\citep{peterson00} or non-existent \citep{done90} X-ray/optical
correlations. In the case of NGC~3516 a strong correlation appeared to
exist in the first set of observations, with the optical leading the
X-rays by 100d \citep{maoz00}, but this correlation was not confirmed
by subsequent observations \citep{maoz02}. However during a 100ksec
\xmm\ observation of NGC~4051\index{NGC~4051}, the UV ($2900\AA$) emission was observed
to lag behind the X-rays by $\sim0.2$d \citep{mason02}. The optical
variability amplitude was much less than that of the X-rays,
consistent with reprocessing.

Our understanding of X-ray/optical variability has, however, improved
considerably recently due largely to the long timescale \xte\ X-ray
monitoring programmes of ourselves and others, coupled with extensive
optical monitoring programmes from groups such as AGNwatch and from
robotic optical telescopes.  In the case of NGC~5548 a strong
X-ray/optical correlation has also been found \citep{uttley035548} but
here the optical band varies at least as much as the X-rays, which is
energetically difficult to achieve with reprocessing. NGC~5548\index{NGC~5548} has a
more massive black hole ($10^{8}$\msun)than NGC~4051 ($10^{6}$\msun) or
NGC~3516\index{NGC~3516} ($4.3 \times 10^{7}$\msun), prompting thoughts that the
correlation might be mass dependent. In higher mass (or lower
accretion rate) black hole systems the disc temperature is lower (for
a given radius, in terms of gravitational radii). Thus the optical
emitting region will be closer to, and will subtend a greater solid
angle at, the central X-ray emitting source. This geometry would
increase the efficiency of reprocessing but would also mean that both
optical and X-ray emitting regions would be subject to more similar
variations than if they were widely separated. Recent X-ray/optical
monitoring of MR~2251-178\index{MR~2251-178} \cite{arevalo08_2251}
another high mass AGN
($10^{8}$\msun), finds a similar pattern to that of NGC~5548.  However
in the case of Mkn~79 \cite{breedt09} ($10^{8}$\msun) we see
correlated X-ray and optical variations on short timescales (tens of
days, Fig.~\ref{fig:79}) but large amplitude changes in the optical
band on $\sim$years timescales which are not seen in the X-ray band.

If we remove the long term trend in the optical light curve of Mkn~79,
we find an extremely strong ($>99.5$\% significance) correlation
between the X-ray and optical variations with a lag very close to zero
days ($\pm \sim2$d). Similar behaviour is seen in other AGN of medium
or low black hole mass (Breedt et al, in preparation). This
correlation provides strong support for the model in which the short
timescale optical variability in AGN is dominated by reprocessing of
X-ray photons. This result is consistent with the observations of
wavelength dependent lags between different optical bands in AGN
\citep{cackett06,sergeev05} which are also best explained by reprocessing.

However the slower optical variations in both MR~2251-178\index{MR~2251-178}
\cite{arevalo08_2251} and Mkn~79\index{Mkn~79} \cite{breedt09} cannot be explained by
reprocessing from an X-ray emission region on an accretion disc both
of constant geometry.  As reprocessing from the accretion disc occurs
on the light travel time which is, at most, a few days, it is
impossible to remove all of the fast variations seen in the X-ray
band.  It might be possible to produce very smooth reprocessed
light curves if the reprocessor is much larger than the accretion disc,
e.g. if it is in the broad line region or torus. Another possibility is
that the slow optical variations can be explained either by varying
the accretion rate in the disc, or by varying the geometry of the
X-ray source or accretion disc. The radius within which half of the
thermal optical emission from the accretion disc in Mkn~79 is produced
is 70$R_{G}$. The viscous timescale at this radius is about 4.7
years. Thus variations in accretion rate, propagating inwards, would
be expected to vary the intrinsic thermal emission from the disc on
that sort of timescale, which is consistent with the optical long term
trends seen in Fig~\ref{fig:79}.  Alternatively, if we approximate the
X-ray emitting corona as a point source at its centroid, above the
accretion disc, we can change the strength of the reprocessed optical
signal by altering the height of that point source. We can also alter
the strength of the reprocessed signal by changing the inner radius of
the accretion disc.  Further work is underway to determine which of
these possibilities is the most likely.

We note that a variable height lamppost model has been invoked to
explain continuum variability coupled with much lower variability of
the iron line in MCG-6-30-15 \cite{miniutti03}. When the source is
close to the disc, a greater fraction of the continuum is
gravitationally bent back to the disc or into the black hole, thereby
reducing the flux as seen by a distant observer but not altering the
reprocessed iron line flux greatly. Although a reasonable explanation
of spectral variability, this model cannot explain the rms-flux
relationship where the observed variability is least at the lowest
continuum fluxes.

\begin{figure}
\includegraphics[angle=270,width=11.5cm]{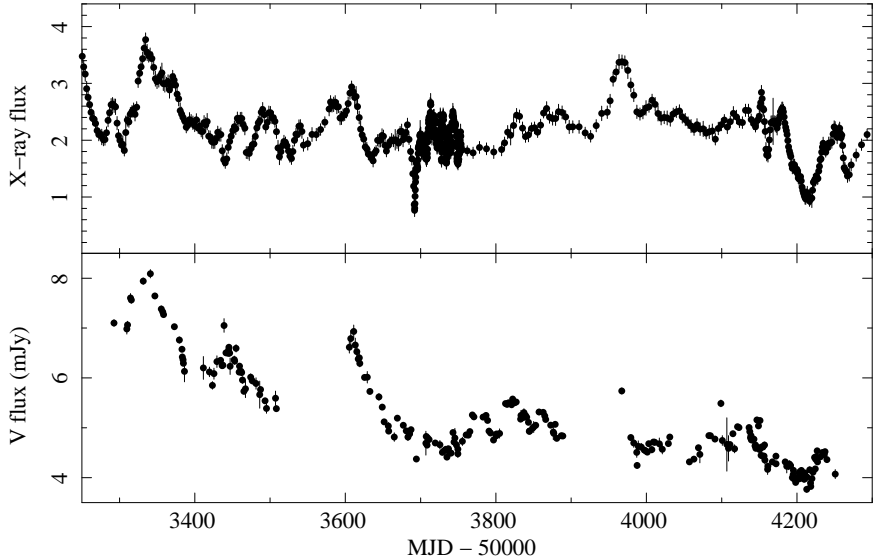}
\caption{Optical and X-ray variations in Mkn~79\index{Mkn~79} \cite{breedt09}. Note
  how the variations are correlated on short timescales (weeks/month),
  probably showing that the optical variations arise from reprocessing
  of the optical emission. However the optical emission varies
  independently on longer timescales, possibly indicating changes in
  the intrinsic emission from the disc due to accretion rate changes
  or possibly indicating changes in the geometry of the X-ray source. }
\label{fig:79}
\end{figure}

\section{Conclusions}

Our understanding of AGN variability has improved enormously in the
last 20 years and the analogy with BHBs is now very strong. Almost all
of the bright AGN with well measured PSDs have PSDs similar to those
of Cyg~X-1 in the soft state. These AGN typically have accretion rates
similar to those of soft state BHBs. Akn~564, which has a very high
accretion rate, perhaps even exceeding the Eddington limit, has a PSD
similar to that of BHBs in the very high state where accretion rates
are similarly high. Ton~S180, another very high accretion rate AGN, has
a less well measured PSD which is also similar to that of a VHS BHB.
As yet we have no firm evidence, from PSD shapes, for a hard-state
AGN, but this result is probably just due to two selection effects.
Hard-state BHBs have low accretion rates ($<$2\% \me) and so have
low X-ray luminosities. Thus, unless they are very nearby, hard
state AGN will be too faint to detect with dedicated monitoring
satellites such as RXTE. Secondly, in order to confirm a hard PSD
state, one must detect the second bend, at low frequency, to a slope of zero
and, as bend frequencies scale inversely with accretion rate, typical
bend timescales will be far too long to be measured within a human
lifetime.

Measurement of the lags between different energy bands as a function
of Fourier timescale provides another strong state diagnostic and in
the very high accretion rate AGN Akn~564 we see a lag spectrum very
similar to that seen in very high accretion rate BHBs. In particular,
we see a rapid change in lag at the frequency at which the PSD changes
from being dominated by one Lorentzian shaped component to another,
again, just as in BHBs.

Characteristic timescales in accreting systems have now been shown to
vary almost exactly linearly with compact object mass and inversely
with accretion rate (in Eddington units). This scaling relationship
applies over a huge range of over 10 decades in timescale. The
physical origin of the high frequency bend in the PSD is not yet
entirely clear but is very likely related to a timescale, probably
viscous, at the inner radius of the optically thick part of the
accretion disc. This radius probably moves inwards with rises in
accretion rate during the hard state but, in the soft state, it may
hit the ISCO where an increase in disc scale height with
increasing accretion rate could explain the decreasing characteristic
timescales.

The fact that the relationship $T_{B} \propto
M_{BH}^{1.12}/\dot{m}_{E}^{0.98}$ is a good fit within the measurement
errors without having to invoke an additional unknown source of error
indicates that no other parameters besides $M_{BH}$ and \me\, have a
major effect on determining \tb. If \tb\, is associated with the inner
edge of the accretion disc then an obvious additional parameter would
be spin, which alters the radius of the last stable orbit and hence of
all associated timescales.  The implication of the good fit is
therefore that the spin of all black holes whose accretion rate is
high enough for them to be powerful X-ray sources is broadly similar.

The discovery of the rms-flux relationship in both AGN and BHBs shows
that the process underlying the variability is multiplicative, rather
than additive. Thus the amplitude of the shorter timescale
variations are modulated by the amplitude of the longer timescale
variations, leading to occasions (e.g. in NGC~4051) where the X-ray flux
almost completely turns off. Standard shot noise models, with
independent shots, are thus ruled out. The rms-flux relationship
predicts a lognormal flux distribution, which is seen in BHBs where
the count rate is high enough for reliable measurement.
The link between variations on different timescales implies that the
light curves will be non-linear, with the greatest non-linearity in the
sources with the highest rms variability, exactly as is seen, e.g. in
the narrow line Seyfert 1 galaxies.

The physical models which best explain the overall patterns of X-ray
variability, i.e. the PSD shape, the lag spectra and the coherence
between different energy bands, are those in which the production of
the variations is separated from the production of the X-rays. Thus
accretion rate variations in the accretion disc
\cite[e.g.][]{lyubarskii97} may propagate inwards and modulate an
X-ray emission region with an energy-stratified emission profile
\cite[e.g.][]{kotov01,churazov01}.

The origin of optical variability in Seyferts, and its relationship to
the X-ray variability, is also partly explained by propagating
fluctuations. Although short-timescale (few days/weeks) optical
variability is well explained by X-ray reprocessing, longer timescale
optical variations, which have no parallel in the X-ray band, may
result from perturbations propagating inwards through the
accretion disc which may take many years to travel from the optical to
X-ray emission region.

There is some evidence that the same general pattern of variability
that is seen in the non-relativistically beamed AGN and BHBs is also
seen in blazars. The X-ray emission mechanism in blazars (eg
synchrotron self-Compton emission) is almost certainly different to
that in Seyfert galaxies (probably thermal Comptonisation) but blazars
also demonstrate the rms-flux relationship and, in at least one
well-observed case (3C~273) the PSD shape is identical to that of
soft-state Seyfert galaxies. Moreover the characteristic PSD timescale
in 3C~273 is entirely consistent with that expected, based on its black
hole mass and accretion rate, from Seyfert
galaxies, without any requirement for modification to take account of
the relativistic motion of the emitting region. The conclusion,
therefore, is that the source of the variations lies outside the jet,
whose emission it modulates, and that that source is probably the same as
in BHBs and Seyfert galaxies.

A major task for future observatories is to extend the variability
studies of Seyfert galaxies to galaxies of very low accretion rate (ie
$<1\%$ \me). The aims are to determine whether such AGN are hard state
systems, whether the transition accretion rate in AGN is the same as,
or different to, that in BHBs and whether such AGN behave in exactly
the same way as scaled-up hard state BHBs. However such AGN are likely
to be faint and the required monitoring timescales will be long. Thus
I encourage the building of a sensitive (sub-mCrab on $\sim$day
timescales) all-sky X-ray monitor, with few arcmin resolution to avoid
confusion problems and with a lifetime of $\sim$10 years rather than
the $\sim$2 years typical of many missions.

\bibliographystyle{plain}

\end{document}